\documentclass[prl,twocolumn]{revtex4}
\usepackage{graphicx}
\usepackage{amssymb,amsmath,bm}
\usepackage{enumerate}

\newcommand{\be}{\begin{eqnarray}}
\newcommand{\ee}{\end{eqnarray}}

\begin{document}

\title{Strong turbulent convection: distributed chaos and large-scale circulation}

\author{A. Bershadskii}

\affiliation{
ICAR, P.O. Box 31155, Jerusalem 91000, Israel
}

\begin{abstract}

Two types of spontaneous breaking of the space translational symmetry in distributed chaos have been considered for turbulent thermal convection at large values of Rayleigh number. First type is related to boundaries and second type is related to appearance of inertial range of scales. The first type is dominated by vorticity correlation integral: $\int_{V} \langle {\boldsymbol \omega} ({\bf x},t) \cdot  {\boldsymbol \omega} ({\bf x} + {\bf r},t) \rangle_{V}  d{\bf r}$ and is characterized by stretched exponential spectrum $\exp-(k/k_{\beta})^{\beta }$ with $\beta =1/2$. The second type is dominated by energy correlation integral: $\int_{V} \langle {\bf u}^2 ({\bf x},t) ~ {\bf u}^2({\bf x} + {\bf r},t) \rangle_{V}  d{\bf r}$ and is characterized by $\beta =3/5$. Good agreement has been established with laboratory experimental data obtained at large values of Rayleigh number $Ra \sim 10^{11}-10^{14}$ (the range relevant to solar photosphere) in upright cylinder cells. Taylor hypothesis transforms the wavenumber spectrum into frequency spectrum $\exp-(f/f_{\beta})^{1/2}$. It is shown that turnover frequency of large-scale circulation (wind): $f_w = f_{\beta}/2$. Results of an experiment in horizontal cylinder are also briefly discussed. The analysis suggests that in this case the large-scale circulation can be considered as a natural (harmonic) part of the distributed chaos.     
\end{abstract}

\maketitle

\section{Spontaneous breaking of space translational symmetry}

  Figure 1 shows a schematic picture of strong turbulent thermal convection in a cell (adapted from Ref. \cite{k}). The hot and cold  plumes generate the large-scale circulation. This picture has been confirmed by direct laboratory observations and numerical simulations. At the
cell center the visual picture should be more 'chaotic' but the physical processes seem to be more complex near the side walls (just because of the strong interaction between the large-scale circulation and the small-scale turbulence). Space homogeneity is hardly expected in such cell.

   The distributed chaos is a basis for turbulence (both homogeneous and inhomogeneous) \cite{b1}. The waves (pulses) driving the distributed chaos have scaling asymptotic
$$
\upsilon (\kappa ) \sim \kappa^{\alpha}     \eqno{(1)}   
$$
for the group velocity $\upsilon (\kappa )$ at $\kappa \rightarrow \infty$. These waves generate the stretched exponential spectrum of the distributed chaos
$$
E(k ) \propto \exp-(k/k_{\beta})^{\beta}  \eqno{(2)}
$$
with
$$
\beta =\frac{2\alpha}{1+2\alpha}   \eqno{(3)}
$$ 
In isotropic homogeneous turbulence the scaling Eq. (1) is dominated by the momentum correlation integral 
$$ 
I_2 = \int  \langle {\bf u} ({\bf x},t) \cdot  {\bf u} ({\bf x} + {\bf r},t) \rangle d{\bf r}  \eqno{(4)}
$$ 
which is related to the space translational symmetry by virtue of the Noether's theorem. 

\begin{figure}
\begin{center}
\includegraphics[width=8cm \vspace{-0.7cm}]{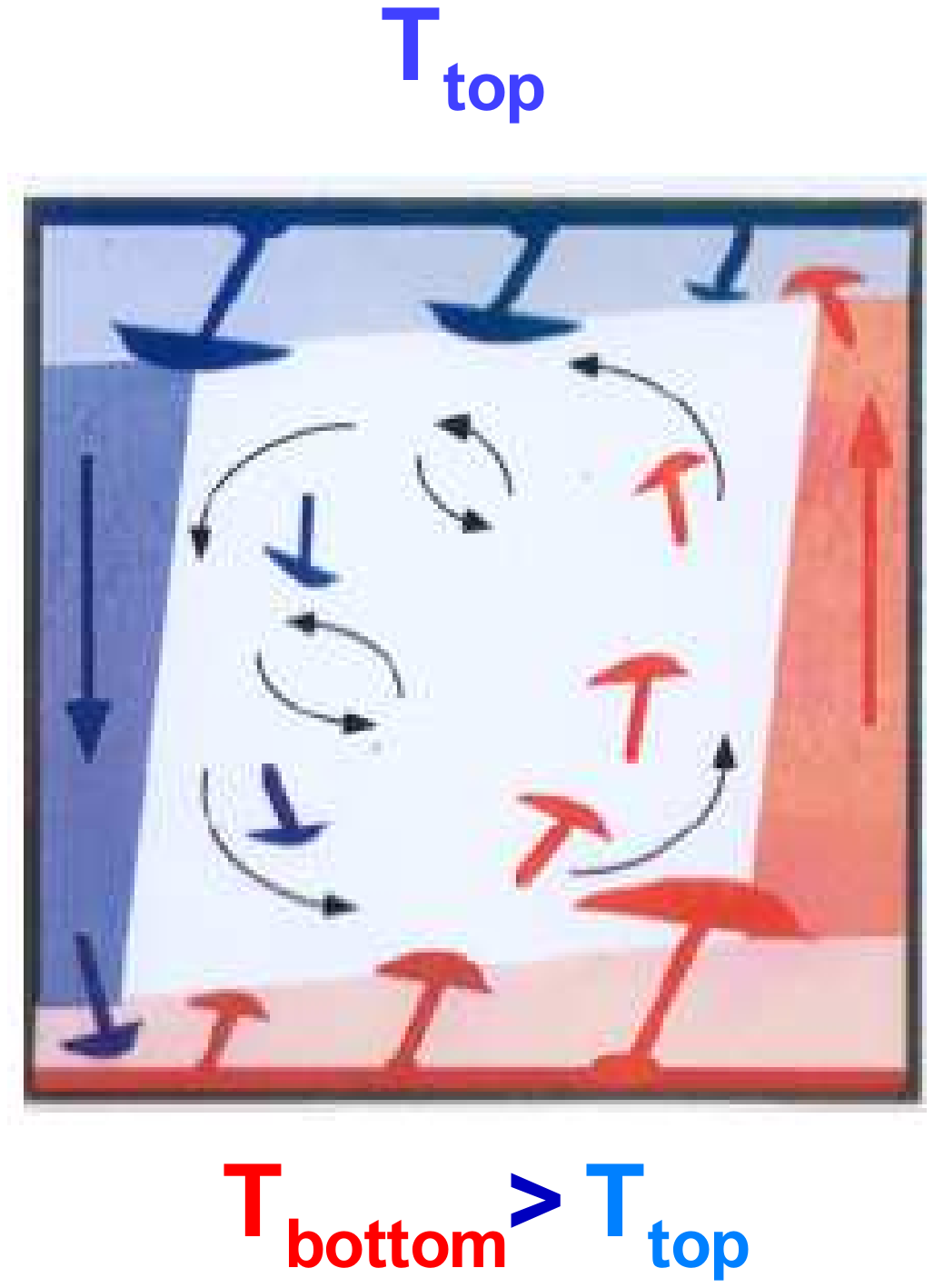}\vspace{-4cm}
\caption{\label{fig1} Kadanoff's cartoon of convective turbulence in a cell (adapted from Ref. \cite{k}.)}
\end{center}
\end{figure}

Substitution of the $I_2$ into Eq. (1) and dimensional considerations result in
$$
\upsilon (\kappa )\propto I_2^{1/2}~\kappa^{3/2} \eqno{(5)}
$$
and, hence, in $\beta =3/4$. 

  Spontaneous breaking of the space translational symmetry (due to space boundaries) switches the control over scaling Eq. (1) to vorticity correlation integral \cite{b2} 
$$
\gamma = \int_{V} \langle {\boldsymbol \omega} ({\bf x},t) \cdot  {\boldsymbol \omega} ({\bf x} + {\bf r},t) \rangle_{V}  d{\bf r} \eqno{(6)}. 
$$ 
In this case substitution of the $\gamma$ into Eq. (1) and dimensional considerations result in 
$$
\upsilon (\kappa ) \simeq a_2~|\gamma|^{1/2}~\kappa^{1/2} \eqno{(7)}
$$
and, hence, in $\beta =1/2$. 

\begin{figure}
\begin{center}
\includegraphics[width=8cm \vspace{-1cm}]{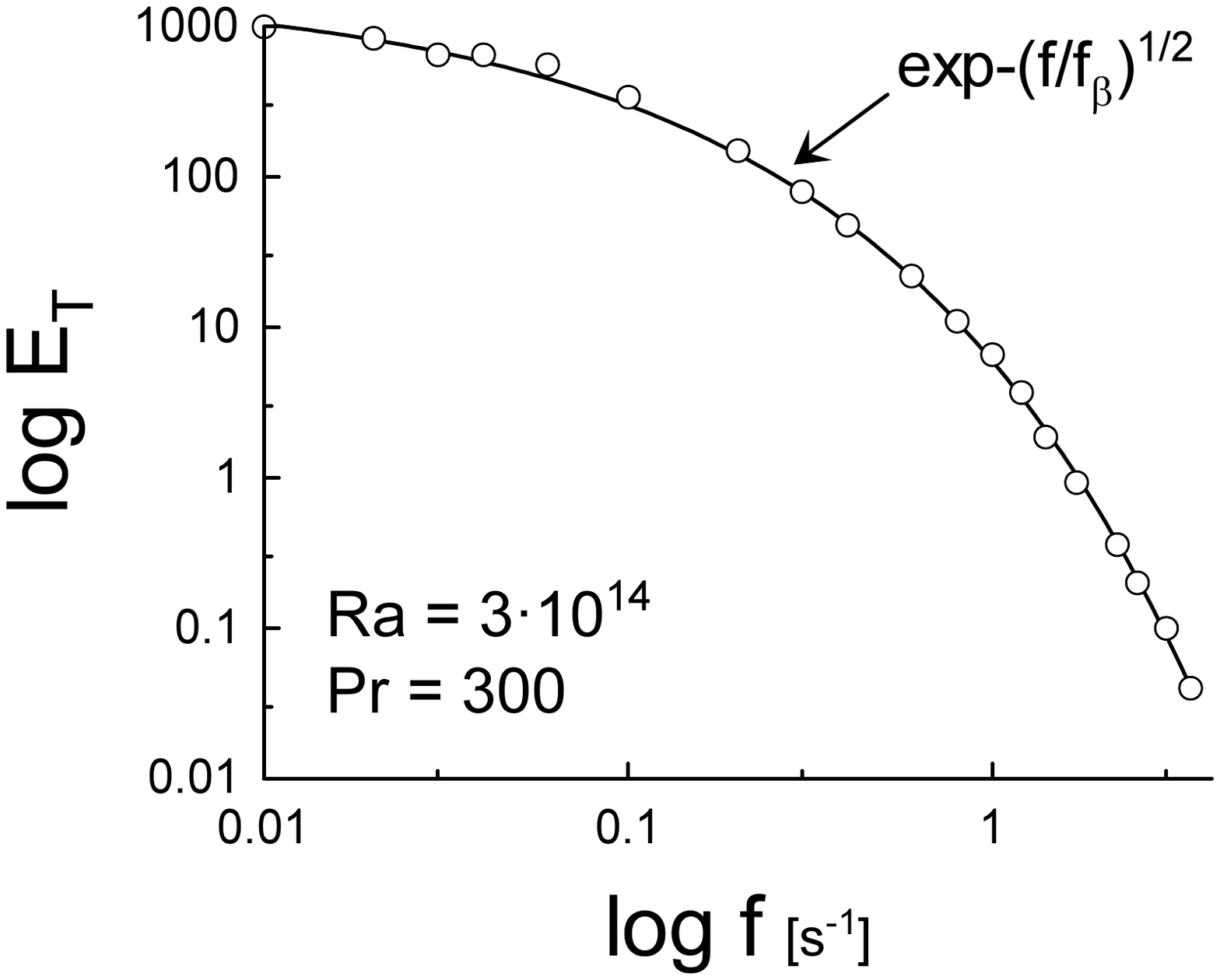}\vspace{-3.5cm}
\caption{\label{fig2} Power spectrum of temperature measured at the cell {\it center} (the data taken from the Ref. \cite{as}). The solid curve indicates the exponential decay Eq. (2) with $\beta =1/2$ (the Taylor hypothesis relating the wavenumber and frequency spectrum has been implied \cite{my},\cite{as}). }. 
\end{center}
\end{figure}

 \begin{figure}
\begin{center}
\includegraphics[width=8cm \vspace{-1cm}]{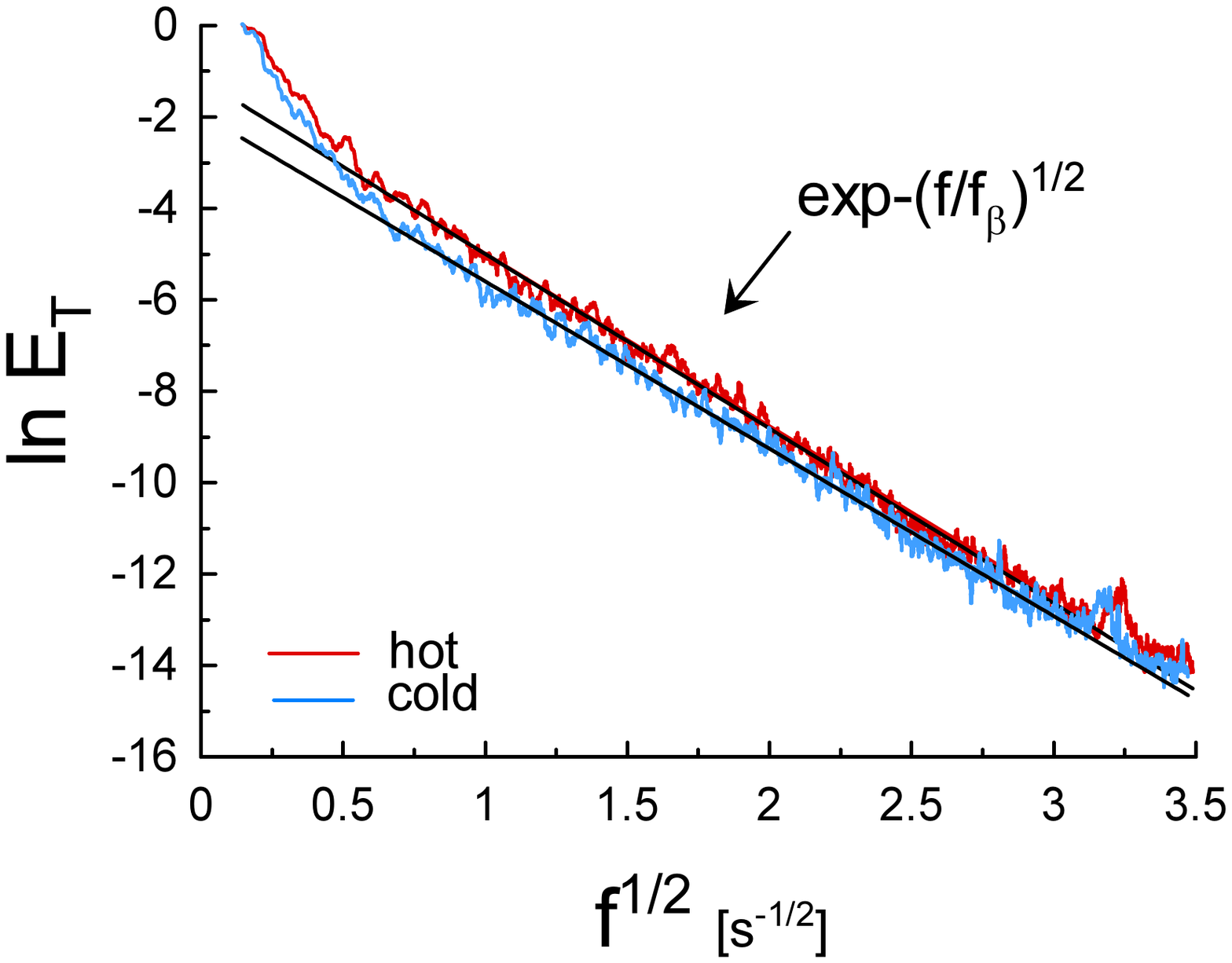}\vspace{-3.8cm}
\caption{\label{fig3} Logarithm of power spectrum of temperature fluctuations as function of $f^{1/2}$.   Measurements were made near side wall at $Ra = 1.5 \cdot 10^{11}$ (the data taken from the experiment described in the Ref. \cite{nssd}). The straight lines indicate the exponential decay Eq. (2) with $\beta =1/2$. }
\end{center}
\end{figure}

  However, there is another (internal) type of spontaneous breaking of the space translational symmetry due to appearance of inertial range of scales nearby the range of scales where the distributed chaos takes place. Since the inertial (Kolmogorov-Obukhov \cite{my}) range of scales is dominated by the energy conservation (a result of the time translational symmetry, due to the same Noether's theorem), this spontaneous symmetry breaking switches the control over the distributed chaos to the energy correlation integral
$$
\mathcal{E}  = \int_V  \langle {\bf u}^2 ({\bf x},t) \cdot  {\bf u}^2 ({\bf x} + {\bf r},t) \rangle_V d{\bf r}  \eqno{(8)}
$$ 
In this case substitution of the $\mathcal{E}$ into Eq. (1) and dimensional considerations result in
$$
\upsilon (\kappa ) \simeq a_2~\mathcal{E}^{1/4}~\kappa^{3/4} \eqno{(9)}
$$
and, hence, in $\beta =3/5$. 

  Stretched exponential spectrum with such value of $\beta$ can be observed even in presumably homogeneous and isotropic turbulence at large Reynolds numbers. \\

\section{Strong turbulent thermal convection}

  Let us return to the strong turbulent thermal convection. Figure 2 shows power spectrum of temperature measured at the cell {\it center} for very large Rayleigh number $3\cdot 10^{14}$ and the Prandtl number $Pr=300$ (experimental measurements reported in the Ref. \cite{as}). 
As it is mentioned in the Ref. \cite{as} the inertial range was completely suppressed by buoyancy in this situation. Therefore, the first (external) type of the spontaneous breaking of the space translational symmetry should prevail here with $\beta = 1/2$. The solid line is drawn in the Fig. 2 in order to indicate the stretched exponential spectrum with $\beta = 1/2$ in a rather large range of scales. The Taylor hypothesis \cite{my},\cite{as} allows to transform the wavenumber spectrum Eq. (2) into the frequency spectrum shown in the Fig. 2.  \\

  As it was mentioned above situation near the side walls of the cell is complicated by the large-scale circulation (wind, cf Fig. 1). Not only the wind interact with the distributed chaos, it also changes direction in a chaotic manner (see, for instance, Refs. \cite{nssd},\cite{sbn},\cite{b2}). If a probe is placed in a fixed location (usually on the horizontal
mid-plane of the cell), then the alterations of the wind direction result in the alterations of the hot (ascending) and the cold (descending) periods in the time series measured by the probe (cf again Fig. 1). Despite the functional form of the spectra can be the same in the hot and cold parts of the wind the parameters can be different. Indeed, Figure 3 shows results of the temperature measurements for the hot and cold parts of the time series (the experiment was reported in the Ref. \cite{nssd}, $Ra = 1.5 \cdot 10^{11}$). One can see that the functional form is the same for these two parts in the distributed chaos range of scales: the stretched exponential with $\beta = 1/2$ (cf. Fig. 2), but the parameters are different: $f_{\beta}\simeq 0.068$ (hot), $f_{\beta}\simeq 0.074$ (cold). 
  
  It is interesting that the averaged value of the $f_{\beta} \simeq 0.071 \simeq 2f_{w}$, where $f_w$ is turnover frequency of the large scale circulation (mean velocity of the wind $\simeq 7~cm/s$ and the perimeter of the apparatus is $200~cm$, cf also Fig. 8 in the Ref. \cite{nssd}). It indicates a strong relation between the distributed chaos and the large-scale circulation. \\
  
  Another type of such relation can be inferred from analysis of the non-splitted (hot and cold mixed) time series. In this case the mixed chaotic attractor should be dominated by properties of the chaotic process of the wind direction alteration. It is shown in the Ref. \cite{b2} that these properties are strongly affected by appearance of the inertial (Kolmogorov) range of scales in the wind velocity fluctuations. Therefore one can expect that the mixed distributed chaos will be dominated by the energy correlation integral, i.e. $\beta=3/5$ in this case (see above). Figure 4 shows the power spectrum of the temperature fluctuations calculated for the full time series (without separation on the hot and cold sub-series). The straight line in this figure indicates the stretched exponential spectrum with the $\beta =3/5$, i.e. the second type of the spontaneous symmetry breaking related to the appearance of the inertial range of scales. \\
  
 \begin{figure}
\begin{center}
\includegraphics[width=8cm \vspace{-0.8cm}]{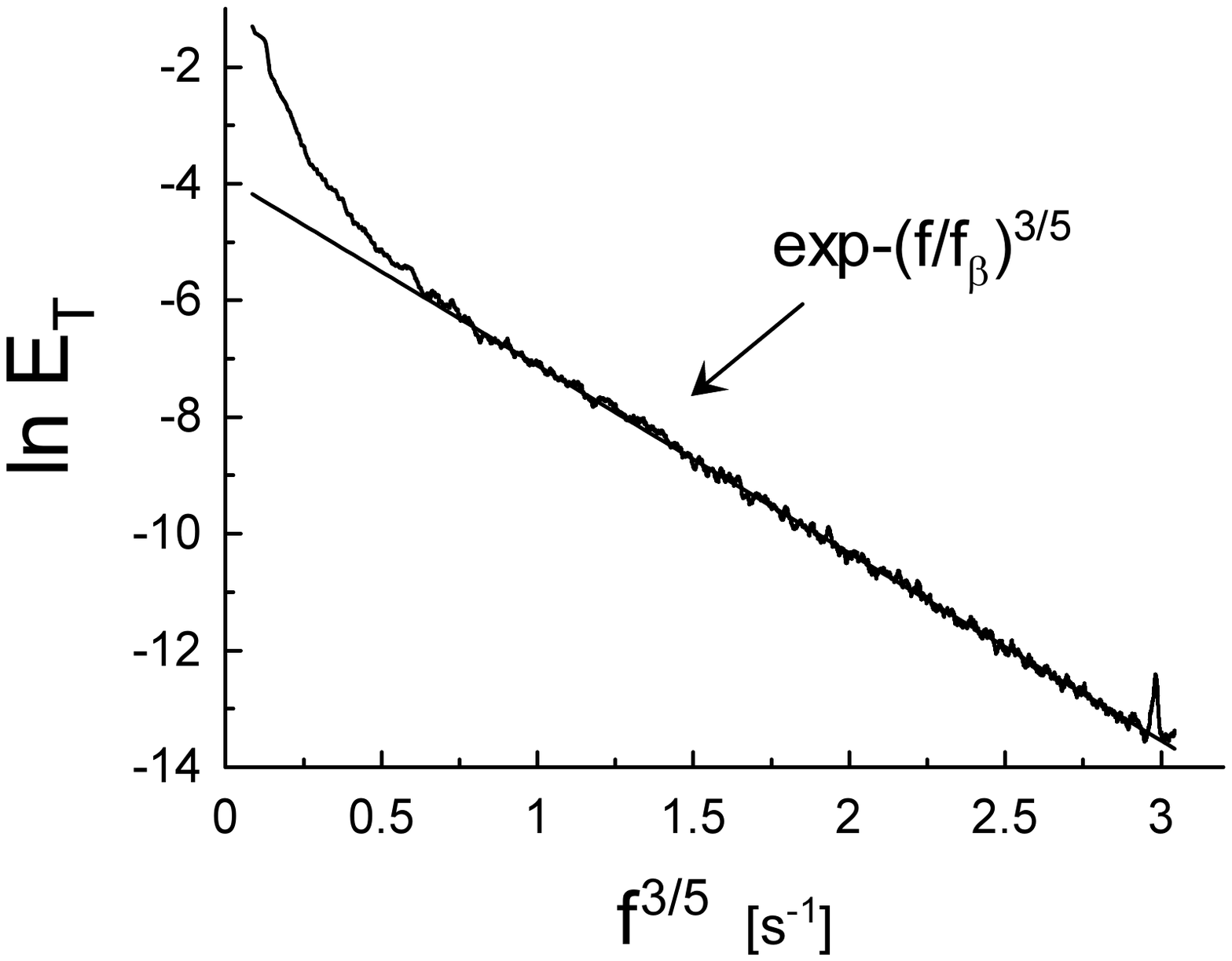}\vspace{-4cm}
\caption{\label{fig4} The same as in Fig. 3 but for the full signal (without separation between the hot and cold periods). The straight line indicates the exponential decay Eq. (2) with $\beta =3/5$.} 
\end{center}
\end{figure} 

\begin{figure}
\begin{center}
\includegraphics[width=10cm \vspace{-2.5cm}]{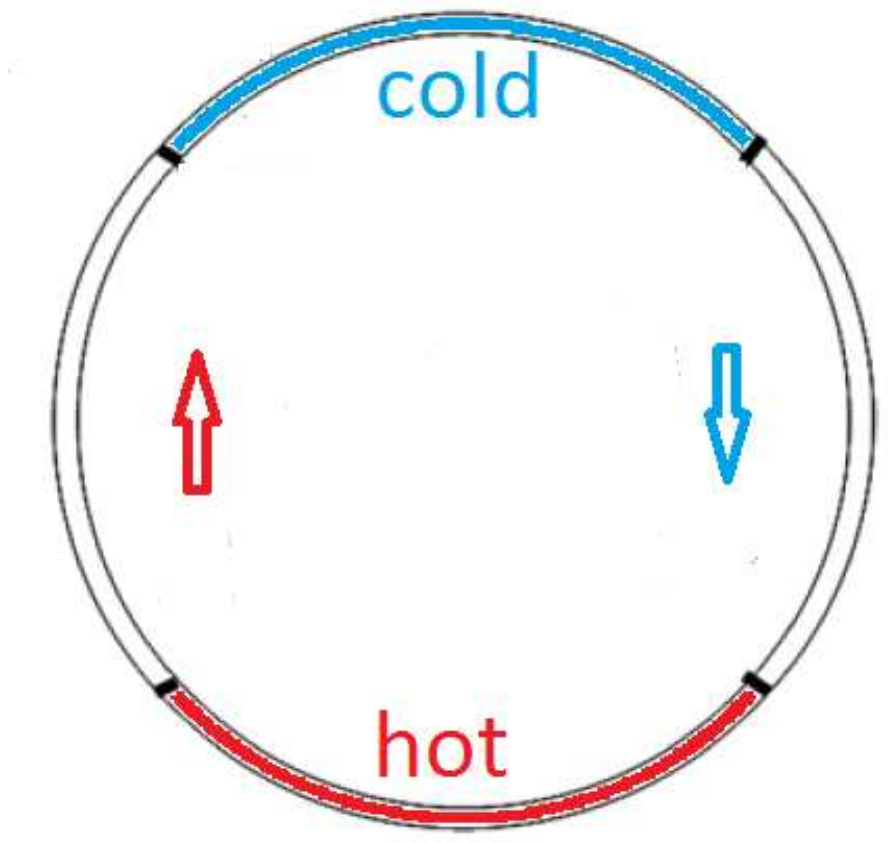}\vspace{-4.5cm}
\caption{\label{fig5} A sketch of the horizontal cylinder apparatus.} 
\end{center}
\end{figure} 

\begin{figure}
\begin{center}
\includegraphics[width=8cm \vspace{-1.1cm}]{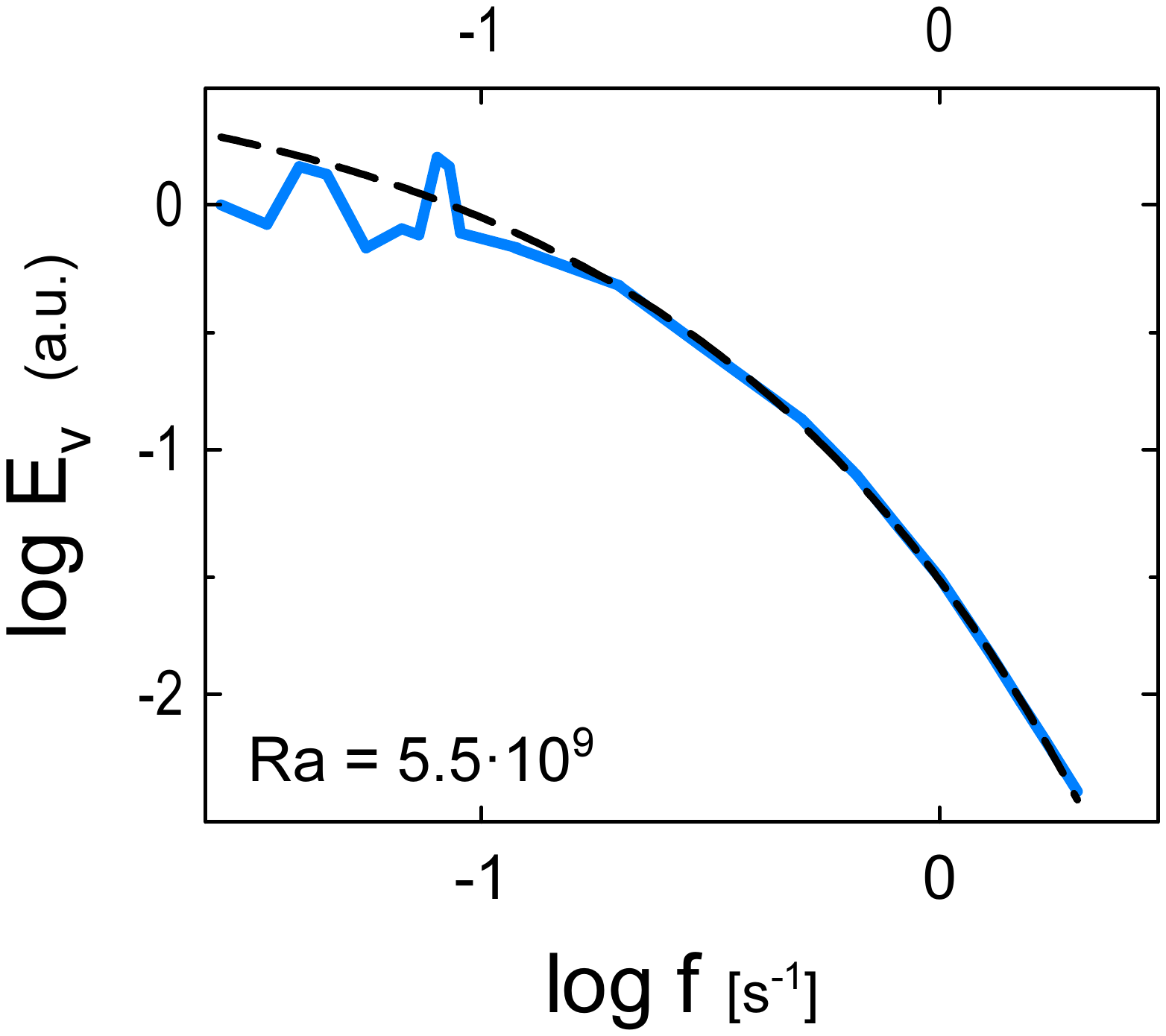}\vspace{-3.4cm}
\caption{\label{fig6} Power spectrum of the horizontal velocity fluctuations for aspect ratio $\Gamma =0.5$ and $Ra=5.5\cdot 10^{9}$ (the data were taken from the Ref. \cite{song}). The dashed curve corresponds to the stretched exponential spectrum Eq. (2) with $\beta =1/2$.}
\end{center}
\end{figure}

\section{Horizontal cylindrical cell}
 
  The above mentioned laboratory experiments used upright cylindrical cells. In a recent experiment, reported in Ref. \cite{song}, a horizontal cylinder was used instead (see a 
sketch of this experimental apparatus in Fig. 5). As we will see relation between the large-scale circulation and distributed chaos in these two cases are different, both qualitatively and quantitatively.  Application of the stretched exponential spectral analysis to the data reported in the Ref. \cite{song} for horizontal velocity spectrum (the dashed curve in Fig. 6) indicates $f_{\beta}$ relation to the periodic switching of the large-scale circulation orientation - the dominant peak in Fig. 6 at $f\simeq 0.08$ Hz, the subharmonic at $f_{\beta}\simeq 0.04$ Hz. Moreover, there is no inertial or Bolgiano range separating between the dominant peak and the distributed chaos range in this case. Actually the distributed chaos range includes the switching frequency. Therefore, the large scale circulation can be considered as a natural (harmonic) part of the distributed chaos in this case (the switching frequency is equal to the large-scale circulation turnover frequency at this value of the aspect ratio). The periodic switching of the large-scale circulation orientation is a result of a spontaneous reflexional symmetry breaking.     
 
\section{Discussion}

 Now we can discuss briefly large-scale inhomogeneity. Patches of different size and temperature pass over a probe located in center of a cell. A full time signal, measured by this probe, will be affected by the above described 'mixing' phenomenon. Depending on conditions this mixed signal will provide an effective value of $\beta$ between 1/2 and 3/5. The first observation of the stretched exponential spectrum in the turbulent thermal convection \cite{wu} reported $\beta \simeq 0.55 \pm 0.05$ for Rayleigh numbers from $7\cdot 10^6$ to $7\cdot 10^{10}$. This large-scale intermittency can affect also results of direct numerical simulations.

   There is also another 'mixing' -  between the two ways of the translational symmetry breaking.  The first one comes from the boundaries, i.e. from the largest scales. Therefore it is stronger in the large-scale part of the distributed chaos range of scales. Final result of this competition depends on concrete conditions in each case. In order to demonstrate this we show in figure 7 the distributed chaos range of scales for temperature power spectrum obtained in an experiment presented in Ref. \cite{hht}. The experiment was performed in an upright cylindrical cell ($Ra = 1.4\cdot 10^{10}$) and the temperature measurements were made near side wall (horizontal mid-plane). The authors of the experiment used a modification of the Taylor hypothesis suggested in Ref. \cite{he}. Normalization of the wavenumber $k$ was made using $\lambda$ - the Taylor microscale. As in the case shown in the Fig. 4 value $\beta = 3/5$ provides the best fit for the full signal data, but the value $\beta = 1/2$ can be also used for a large-scale part of the distributed chaos range, as one can see from the Fig. 7. Moreover, $k_{\beta}$ extracted from the Fig. 7 corresponds to the turnover frequency of the large scale circulation in this case. \\
 
  \begin{figure}
\begin{center}
\includegraphics[width=8cm \vspace{-0.8cm}]{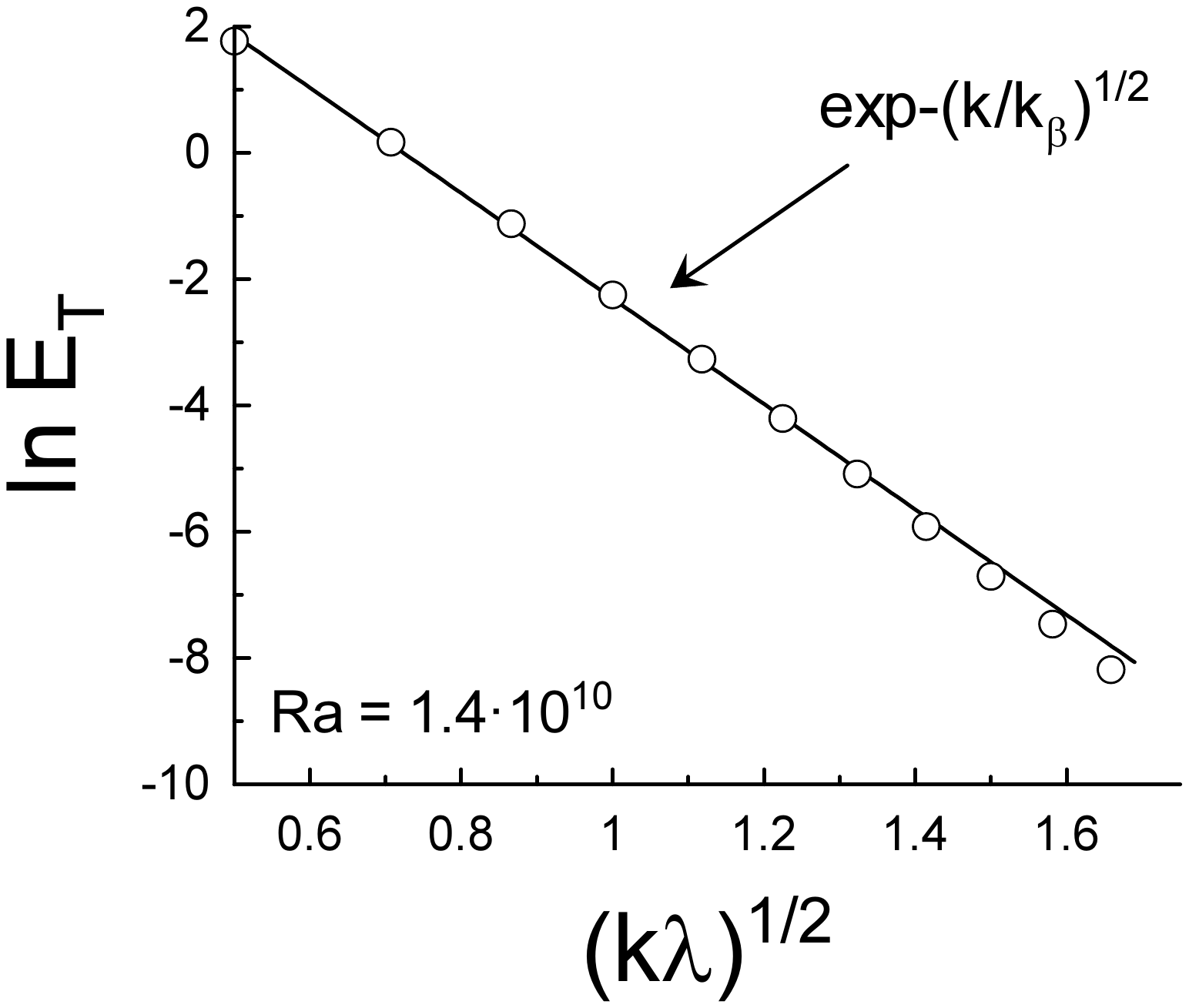}\vspace{-4cm}
\caption{\label{fig7} Temperature power spectrum for the full signal (without separation between the hot and cold periods). The straight line indicates the exponential decay Eq. (2) with $\beta =1/2$. The data were taken from Ref. \cite{hht}.} 
\end{center}
\end{figure}
 
 Finally it should be noted that the range $Ra \sim 10^{11}-10^{14}$ is relevant to turbulent convection in solar photosphere, where role of a boundary (causing the inhomogeneity) plays the $\tau = 1$ optical depth surface \cite{rin},\cite{b3}.

\section{Acknowledgement}

I thank K. P. Iyer, J. J. Niemela, K. R. Sreenivasan and V. Steinberg for sharing their data and discussions.

\end{document}